\documentclass[a4paper]{jpconf}
\usepackage{graphicx}
\usepackage{slashed}
\usepackage{amsmath, amsthm, amssymb}
\begin{document}
\title{Vacuum Cherenkov radiation and photon decay rates from generic Lorentz Invariance Violation}

\author{H. Mart\'inez-Huerta and A. P\'erez-Lorenzana}

\ead{hmartinez@fis.cinvestav.mx and aplorenz@fis.cinvestav.mx}

\address{Physics Department, Centro de Investigacion y de Estudios Avanzados del IPN,\\
Apartado Postal 14-740, 07000, Mexico City, Mexico.}

\begin{abstract}

Among the most studied approaches  to introduce the breaking of Lorentz symmetry, the generic approach is one of the most  frequently used for phenomenology, it converges on the modification of the free particle  dispersion relation. Using this approach in the photon sector, we have calculated the squared probability amplitude for vacuum Cherenkov radiation and photon decay by correcting the QED coupling at tree level and first order in LIV parameters. For the lower order energy correction we calculate the emission and decay rate for each process.

\end{abstract}

\section{Introduction}

Cosmic rays are the  most energetic phenomena known so far, reaching energies of several decades of EeV \cite{XMAX,GZK}. Among the many particulars their study and understanding reveal, they also provide an energy window to test fundamental physics. Such is the case of the search for signatures of Lorentz Invariance Violation (LIV), mainly motivated by Quantum Gravity theories and string theories  \cite{QG1,QG2, QG3, QG4}. Due the the nature of this symmetry, the derived physics from LIV tends to be unique and energy dependent \cite{DIS2, DIS1}. Therefore, its consequences at the highest energies and  very long distances could be identified in the current observatories and experiments.

Following this spirit, in this paper we present a generic approach and a phenomenological first order correction to the production rates of two processes that could have a significant impact on cosmic particle propagation. 
The generic mechanism for introducing LIV, often found in the literature \cite{DIS2,DIS1, DIS3,DIS4, GUNTER-PH, VCR, GUNTER-PD, LIV-proc}, is summarized in an explicit not Lorentz invariant (LI) term added to the free particle Lagrangian density that will converge into the following generic correction to the dispersion relation:
\begin{equation}\label{eq_S}
    S_{a} = E_a^2 - p_a^2 = m_a^2 \pm \alpha_{a,n}A^{n+2},
\end{equation}
where $E_a$ and $p_a$ stand for the four-momenta  associated with an $a$ particle species. Additionally, for particular models, $A$ can take the form of $E$ or $p$, however, for the ultra relativistic limit where $m_a\ll\{E, p \} $, any particular choice of $A$ will be equivalent. The coefficient $\alpha_{a,n}$ in Eq.~(\ref{eq_S}), parametrizes the particle species dependent LIV correction, where $n$ expresses the correction order to the mass shell. It is common to associate a generic $\alpha_{n} \approx E_{QG}^{-n}$, where $E_{QG}$ is the scale of Quantum Gravity or the scale of the expected new physics beneath. Several methods are used in the search for LIV signals, some of them can lead to lower limits to $E_{QG}$ \cite{HESS-LIV,FERMI-LIV,GRB-LIV, HAWC-LIV}, which is expected to be close to  $10 ^ {19}$~GeV~.


In the next section,  we have applied the generic correction shown in Eq.~(\ref{eq_S}) for photons to derived the LIV corrected square amplitude at tree level from the diagrams in figure 1. It is worth stressing that such processes are forbidden in the standard theory by energy-momentum conservation, but under the LIV hypothesis they can be possible. Both of them are used in the search for LIV evidence. The first one is the emission of a single photon by a charged particle that propagates in vacuum  and it is frequently named  vacuum Cherenkov radiation \cite{GUNTER-PH,VCR}. The second is LIV photon decay \cite{GUNTER-PH, GUNTER-PD}, and it is motivated by the LIV extra term in Eq.~(1) that can be read as an effective photon mass and it will depend on the LIV coefficients; the simplest process will produce an electron - positron pair. 
Once the corrected square amplitude has been derived, we used it to find the process rates. 

\section{LIV corrected rates}

\begin{figure}[h]
	\begin{minipage}{14pc}
	    {\centering
		\includegraphics[width=9pc]{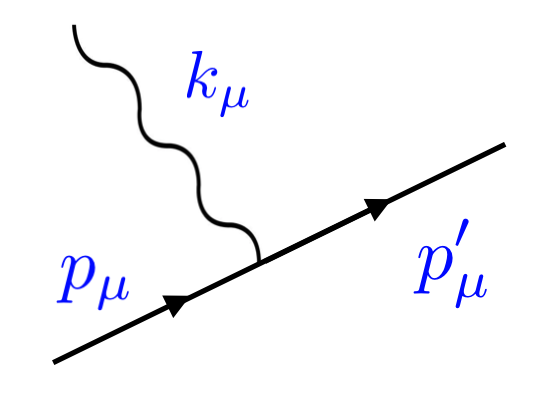}
		\caption{\label{fig1}Diagram for LIV vacuum Cherenkov radiation.}}
	\end{minipage}\hspace{8pc}%
	\begin{minipage}{14pc}
	    {\centering
		\includegraphics[width=11pc]{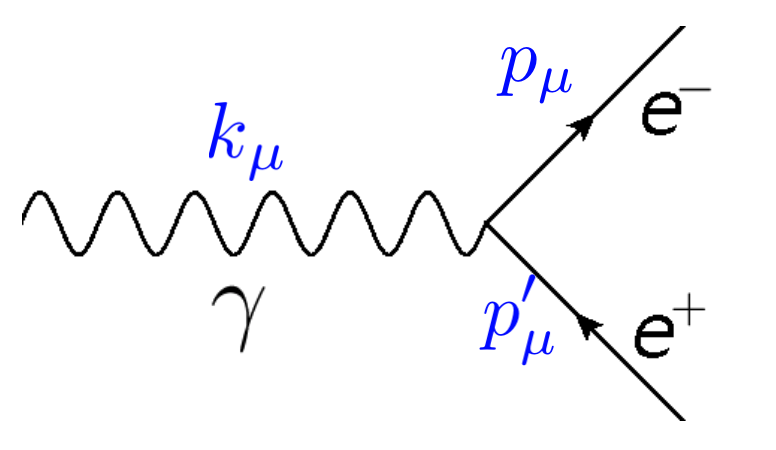}
		\caption{\label{fig2}Diagram for LIV photon decay.}}
	\end{minipage} 
\end{figure}

Following the diagrams in figures 1 and 2, let be $(E_{p},p)$ and $(E_{p'},p')$  the four-momenta for the initial and final particle in the particular case in figure 1, but the four-momenta for the $e^+$-$e^-$ pair in figure 2, as indicated.  We will call $(\omega,k)$ the photon four-momenta in both the cases. With such a consideration, we can express the square probability amplitude for both processes with a single expression, given by
\begin{equation}\label{eq_sum}
\begin{aligned}
    \frac{1}{2}\sum_{spin} |M|^2 & = 
    (\slashed{p} - \slashed{k} + m)(e^2 \gamma^\mu \gamma^\nu) (\slashed{p}+m)\epsilon_\mu\bar{\epsilon}_\nu \\
    & = 2 \sum_{spin} (p^\mu p^\nu - k^\mu p^\nu - p^\nu p^\mu + k^\nu p^\mu + p^\alpha p_\alpha g^{\mu\nu} - k^\alpha k_\alpha g^{\mu\nu} +m^2 g^{\mu\nu} )\epsilon_\mu\bar{\epsilon}_\nu~,
\end{aligned}
\end{equation}
where
Minkowski metric $g^{\mu\nu}=\eta^{\mu\nu}$  and Dirac gamma matrices standard properties under a Lorentz Invariant (LI) theory are assumed. Since the correction is taken at first LIV order correction,  any term 
$\geq \alpha_{n}^2$ is negligible and the Ward identity is preserved. 

From Eq.~(\ref{eq_S}), $S_\gamma=k^\alpha k_\alpha = \alpha_n k^{n+2}$, and assuming the charged particle dispersion relation to be Lorentz conserving along the process, we note that
\begin{equation}
	S_{p} = S_{p'}= (p - k)^\alpha(p - k)_\alpha = S_p + S_\gamma -2k^\alpha p_\alpha,
\end{equation}
then
\begin{equation}
	2k^\alpha p_\alpha = S_\gamma = \alpha_{n} k^{n+2},
\end{equation}
where we have dropped out the particle index on the photon LIV parameter.
Hence, Eq. (\ref{eq_sum}) becomes:
\begin{equation}
	\frac{1}{2}\sum_{spin} |M|^2 = e^2 |4m_a^2 - \alpha_{n} k^{n+2} | 
\end{equation}
The absolute value is taking in order to ensure physical congruence due the limits and the generality in the sign of $ \alpha_{n}$.

Once the modified squared amplitude for both process is found,  we have derived the modified decay and emission rates on an arbitrary preferential frame and for $n=1$. 

For vacuum Cherenkov radiation we have found that 
\begin{equation}\label{eq_VCR_gamma}
	\Gamma_{a\rightarrow a\gamma} =   \dfrac{e^2}{4\pi} \dfrac{1}{4E_a} 
 	\int_0^\pi \frac{|4m_a^2 - \alpha_1k_+^3| }{\omega(k_+,\alpha_{1} )} 
	   \frac{k_+^2 \sin\theta d\theta}{| p_a \cos\theta -k_+  - \frac{(1+\frac{3}{2}\alpha_1k_+)}{\sqrt{1+\alpha_{(1)}k_+}} \sqrt{k_+^2 + E_a^2-2k_+p_a\cos\theta}|},
\end{equation}
where the photon momenta modes from the corrected energy-momenta conservation are 
\begin{equation}\label{eq_kplus}
	k_\pm =  \frac{1}{2p_a\cos\theta}
	\left( E_a^2   \pm \sqrt{E_a^4 +  \frac{4p_a\cos\theta}{\alpha_{1}}(E^2_a - p_a^2\cos^2\theta)}\right).
\end{equation}
We have used only $k_+$, since we are looking for a phenomenological approach and $k_-$ will not be physical.   

On the other hand, for LIV photon decay we have: 
\begin{equation}\label{eq_PD_gamma}
	\begin{aligned}
	\Gamma_{\gamma\rightarrow e^+e^-} =  &  \dfrac{e^2}{4\pi}  \frac{|4m_e^2 - \alpha_{1}k^{3}| }{4 \omega(k,\alpha_{1})} 
	& \int_0^{\pi} \sum_{p=p_\pm} \frac{p^2\sin\theta d\theta }{| (k\cos\theta - p)E_e - p\sqrt{k^2 + E_e^2-2kp\cos\theta}  | }, 
	\end{aligned}
\end{equation}
where the momenta modes from the corrected energy-momenta conservation are
\begin{equation}\label{disc}
	p_{\pm}= \frac{1}{2(\alpha_1k + \sin^2\theta)}\left( \alpha_1 k^{2}\cos\theta  
	 \pm \sqrt{\alpha_1^2 k^{4}\cos^2\theta-4(\sin^2\theta+\alpha_1 k)(1+\alpha_1k)m_e^2} \right) .
\end{equation}
Above, $e$ stands for the electron charge, $\theta$ is the angle between final particles,  $p_a=\sqrt{E_a^2-m_a^2}$ for the given particle $a$ and $\omega (k,\alpha_1) = |k|\sqrt{1+\alpha_{1}k}$. 
Notice that the formulae for photon decay are actually generic for any fermion pair in the final state, provided the corresponding mass is used.
There exist vacuum Cherenkov radiation and photon decay rates obtained from different 
LIV approaches, for instance, from the 
minimal Standard-Model extension by the spontaneous breaking of Lorentz symmetry
\cite{SME} and from the introduction of Lorentz violating operators of dimensions four and six can be found in \cite{TWO-side, CROSS}. 

The last integration in eqs. (\ref{eq_VCR_gamma}) and (\ref{eq_PD_gamma}) are numerically performed for both expressions. For $n=1$, the results are depicted in figures 3 and 4. Since we are looking for an approach at the highest energies, we choose $m_{a}= m_{proton}$ in the first considered process due its relevance for cosmic rays \cite{XMAX,MIXED2}. A small angle approach is taken for vacuum Cherenkov radiation in order to clarify the threshold and in LIV photon decay to allow the process in a large energetic window. 

\begin{figure}[h]
	\begin{minipage}{17pc}
	    {\centering
		\includegraphics[width=18pc]{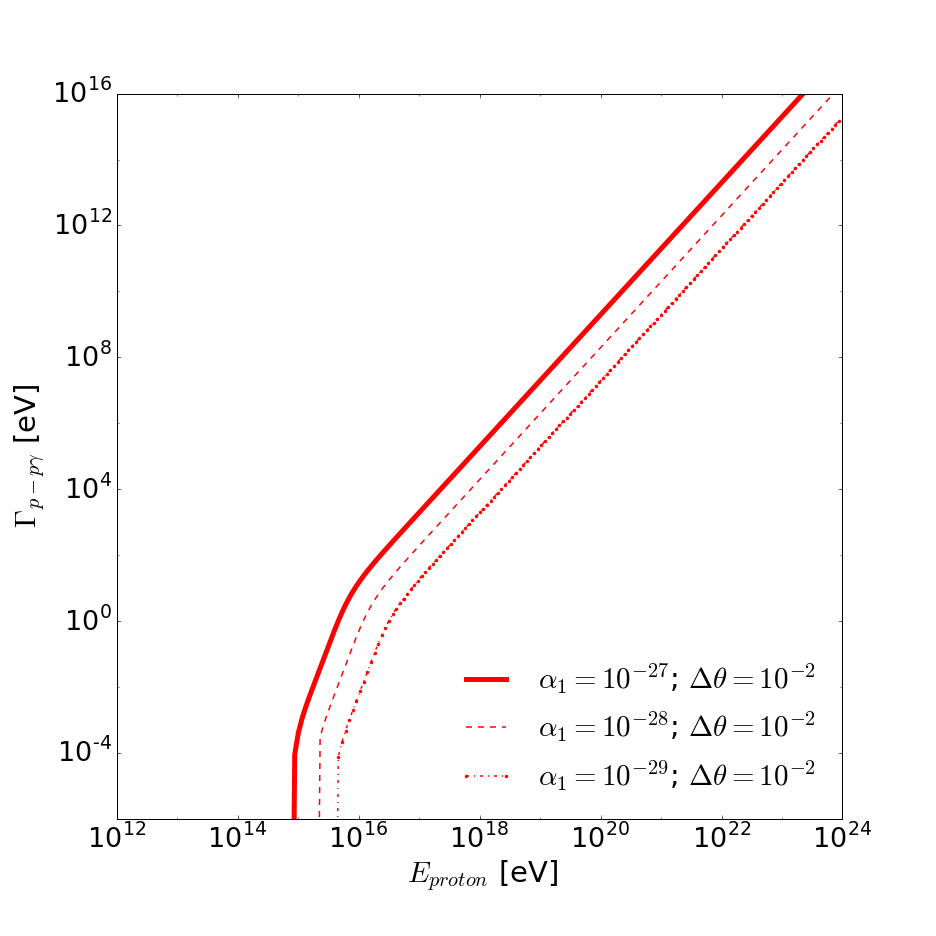}
		\caption{\label{fig3}Emission rate for LIV vacuum Cherenkov radiation for n=1, and $m_{a} = m_{proton}$. Lower values of the LIV photon parameter 	lead to a higher energetic threshold. }}
	\end{minipage}\hspace{3pc}%
	\begin{minipage}{17pc}
	    {\centering
		\includegraphics[width=18pc]{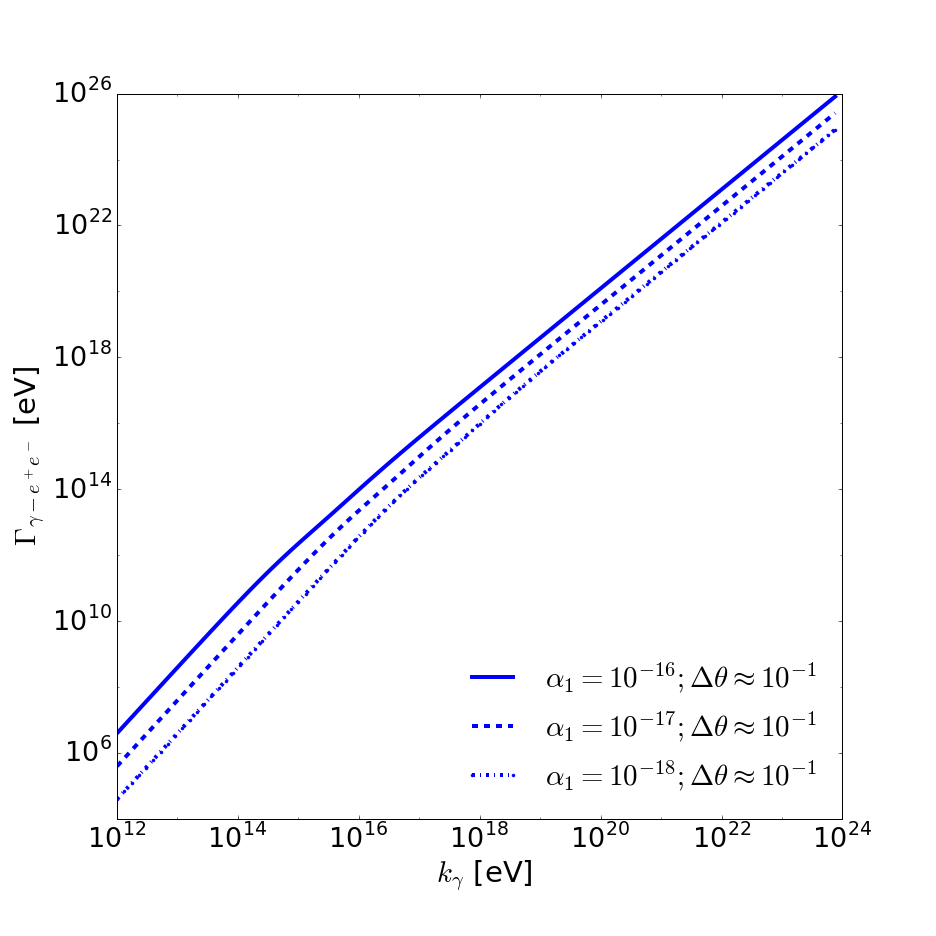}
		\caption{\label{fig4}Decay rate for LIV photon decay into electron positron pairs, for n=1. Lower values of the LIV photon parameter lead to a lower decay rate. \\}}
	\end{minipage} 
\end{figure}

As it can be seen, the phenomena are sensitive to the LIV term. For vacuum Cherenkov emission, there is an energy threshold, proportional to ($E_{a},m_{a},\alpha_{1}$), that preserves unaffected the physics below it. The threshold grows inversely with the value of the LIV photon parameter, $\alpha_{1}$, and above it,
the emission rate grows with charged particle energy. 
In the photon decay case, the rate decreases with the LIV photon parameter, $\alpha_{(1)}$, and grows with the photon momenta. 
There is also a threshold that turns off photon decay 
process at low energies that comes from 
energy-momentum conservation in the LI fermion sector and the corrected LIV photon.

\section{Conclusions}
We have found a generic first order LIV correction to the emission and decay rates for vacuum Cherenkov radiation and LIV photon decay. Both processes, kinetically forbidden in a LI theory, can be possible under LIV hypothesis. The possible consequences of both processes decrease while the LIV photon coefficient $\alpha_{1} \rightarrow 0$. Both processes can lead to different expected physics at the most energetic scenarios, such as cosmic rays, but we will present such an analysis in a future work.

\ack
This work was partially supported by Conacyt grant No. 237004.

\section*{References}

\bibliography{bibfile}

\end{document}